\documentclass[twocolumn,prb,showkeys,showpacs,floatfix]{revtex4}

\usepackage{graphicx}
\usepackage{dcolumn}
\usepackage{bm}
\usepackage{makeidx}
\usepackage{amsmath}
\usepackage{amsthm}
\usepackage{amssymb}
\usepackage{amsxtra}
\usepackage{amscd}
\usepackage[mathscr]{eucal}
\usepackage{boxedminipage}
\usepackage{bbm}
\usepackage{epsfig}
\usepackage{epic,eepic}
\usepackage{latexsym}

\newcommand{\be}{\begin{equation}}
\newcommand{\ee}{\end{equation}}
\newcommand{\bea}{\begin{eqnarray}}
\newcommand{\eea}{\end{eqnarray}}
\newcommand{\Eq}[1]{Eq.~(\ref{#1})}
\newcommand{\Fig}[1]{Fig.~\ref{#1}} 
\newcommand{\olcite}[1]{[\onlinecite{#1}]}  

\newcommand{\tdec}{t_\mathrm{dec}}
\newcommand{\tmix}{t_\mathrm{mix}}
\newcommand{\Pint}{P_\mathrm{int}}
\newcommand{\Pcl}{P_\mathrm{cl}}
\newcommand{\tdFLO}{\tau^D_\mathrm{FLO}}
\newcommand{\aFLO}{a_\mathrm{FLO}}

\newcommand{\lth}{\lambda_\mathrm{th}}
\newcommand{\rhored}{\tilde{\rho}}
\newcommand{\rhoth}{\rho_\mathrm{th}}
\newcommand{\Dpath}[3]{\int_{#2}^{#3} {\cal{D}} \left[ {#1} \right]}
\newcommand{\xbar}{\bar{x}}
\newcommand{\Qbar}{\bar{Q}}

\begin{document}


\title{Decoherence without dissipation?}


\author{Dominique Gobert}
 \email{gobert@lmu.de}
\author{Jan von Delft}
 \affiliation{Sektion Physik und CeNS,
 Ludwigs-Maximilians-Universit\"at, Theresienstr.~37, 80333
 M\"unchen, Germany}
\author{Vinay Ambegaokar}
 \affiliation{Laboratory of Atomic and Solid State Physics, Cornell
 University, Ithaca, New York 14853, USA}

\date{\today}


\begin{abstract}
In a recent article,\cite{FordOConnell01} Ford, Lewis and O'Connell
discuss a thought experiment in which a Brownian particle is subjected
to a double-slit measurement.
Analyzing the decay of the emerging interference pattern, they derive a
decoherence rate that is much faster than previous results and even
persists in the limit of vanishing dissipation.
This result is based on the definition of a certain attenuation factor,
which they analyze for short times.
In this note, we point out that this attenuation factor
captures the physics of decoherence only for times larger than a
certain time $\tmix$, which is the time it takes until the two
emerging wave packets begin to overlap. 
Therefore, the strategy of Ford et al of extracting the decoherence time from the regime
$t<\tmix$ is in our opinion not meaningful.
If one analyzes the attenuation factor for $t > \tmix$, one
recovers familiar behaviour for the decoherence time; in particular, 
no decoherence is seen in the absence of dissipation.
The latter conclusion is confirmed with a simple calculation of the
off-diagonal elements of the reduced density matrix.
\end{abstract}

\maketitle

\section{Introduction}

It is widely accepted that a rapid loss of coherence
caused by the coupling to environmental degrees of freedom is at the root
of the non-observation of superpositions of macroscopically distinct
quantum states.  There is a now well established theoretical 
scheme---``Dissipative Quantum Mechanics''---for studying the details of
this phenomenon, and for analyzing its time scale.\cite{Weiss93} 
It has also become possible to observe decoherence in a  
variety of experiments in mesoscopic physics 
\cite{ChiorescuMooij03,NakamuraTsai99} and quantum optics. 
\cite{KokorowskiPritchard01,ArndtZeilinger99}

In a recent publication,\cite{FordOConnell01} Ford, Lewis and O'Connell
(henceforth abbreviated as FLO) discuss a thought 
experiment in which a Brownian particle initially in thermal 
equilibrium with its environment is subjected to a double-slit
position measurement, giving rise to an interference pattern.
Analyzing the decay of this pattern, they derive a
decoherence time that is much shorter than suggested by previous
calculation.\cite{Venugopalan99}  They suggest the tentative explanation that
inital particle-bath correlations, which drastically alter the short-time
behaviour of the Brownian particle, were not properly taken into
account in previous work.
Because the decoherence time calculated by FLO remains finite even
in the absence of any coupling to the environment, they describe their
result as decoherence without dissipation.\cite{FordOConnell01b}

This is very puzzling.  The usual physical picture of
decoherence\cite{Weiss93,ambeg93} is that averaging over unobserved degrees of
freedom (the ``environment'') leads to non-unitary time evolution, with a
consequent loss of information.  If there is no coupling to the
environment, there will be no such loss. This picture agrees with another
commonly accepted definition of decoherence,
as the decay of the off-diagonal elements of the reduced density
matrix (more precisely, the decay of the interference part
$\rho_\mathrm{int}$ to be defined in section \ref{OD}). 
Without environmental coupling, the time evolution of the
system -- and thus of $\rho_\mathrm{int}$ -- is unitary, the norm of
$\rho_\mathrm{int}$ is constant and does not decay.

In light of these obvious remarks, it is interesting to ask what FLO
mean by decoherence.
In this type of double slit experiment it is essential that the two
initially separated parts of the wave function eventually overlap if
an interference pattern in the probability density
of the particle is to be observed.\cite{CaldeiraLeggett85}
In the thought experiment considered by FLO, this overlap becomes
sizeable only after the broadening of the two wave packets emerging
from either slit becomes equal to their initial separation, which
happens after a certain timescale $\tmix$, to
be defined in \Eq{tmix_def} below.
On much shorter time scales, 
the interference pattern is not only influenced by
the presence (or absence) of coherence, but also by the degree of
overlap of the wave functions,
which makes it difficult, if not impossible, to extract from the
probability density alone a quantitative
measure of decoherence that is meaningful for $t < \tmix$.
As will be shown, the decoherence time obtained by FLO is
much shorter than $\tmix$ and hence merely reflects an arbitrariness in the
definition of decoherence at these short time scales.

The outline of this paper is as follows: 
In the following section, we give a brief summary of FLO's work and
present, in section \ref{P_result}, the results of an alternative 
derivation (given in appendix \ref{Grabert})
of the probability density of the particle, which 
fully agrees with that of FLO.
Section \ref{critique} demonstrates that dynamic effects (i.e.~the
spreading of the wave packets), rather than actual decoherence effects, enter
FLO's definition of decoherence for times shorter than $\tmix$. 
This fact is further illustrated in section \ref{alternative} in which
a definition of decoherence is given that incorporates the wave
packet spreading in a different (and equally arbitrary) way, but 
gives rise to an entirely different picture on these short time
scales.
Finally, in section \ref{OD}, the off-diagonal elements of the reduced density
matrix, which allow the definition of a decoherence measure valid also for
$t < \tmix$, are analyzed, and no decoherence without dissipation is found.

\section{Summary of Ford, Lewis and O'Connell's results}

In the thought experiment discussed by FLO,\cite{FordOConnell01} a
one-dimensional free Brownian particle, in thermal 
equilibrium with its environment, is suddenly (at time $t_1=0$, say) subjected
to a double-slit position measurement.
This measurement is described as a weighted sum of
projectors 
$P = \int dx \alpha(x) |x\rangle \langle x|$
acting from the left and right on the density
matrix, such that after the measurement, the state is described as
\bea
&&
\rho_\mathrm{ini} (x,x', \{Q_\alpha\}, \{Q_\alpha'\}) \nonumber \\
&=& \alpha^*(x) \alpha(x') \rho_\mathrm{th}  (x,x', \{Q_\alpha\}, \{Q_\alpha'\}),
\eea
where $\rho_\mathrm{th}(x,x', \{Q_\alpha\}, \{Q_\alpha'\})$ denotes the density matrix of
a particle (described by the coordinate $x$) in thermal equilibrium
with its environment (described by a set of coordinates $Q_\alpha$).
The measurement function $\alpha$  describes the transmittance of the
double slit and is taken as a sum of two Gaussian functions with width
$2 \sigma$, and separated by a distance $d \gg \sigma$:
\be
\label{alpha_2slit}
\alpha(x) = \frac{N^{1/2}}{(8 \pi \sigma^2)^{1/4}} \cdot 
  \left( e^{ - \frac{(x - d/2)^2}{4 \sigma^2} } 
+ e^{ - \frac{(x + d/2)^2}{4 \sigma^2} } \right),
\ee
$N = (1 + e^{- d^2 / (8 \sigma^2) } )^{-1}$ being a normalization
constant, such that $\int dx |\alpha(x)|^2 = 1$. 
The particle dynamics are calculated in the framework of a quantum
Langevin equation,\cite{FordKac87, FordLewis86} which 
describes the dynamics of a particle coupled to a dissipative
environment with Ohmic characteristics.

Within this framework, FLO calculate the probability density $P(x,t) =
\rhored(x,x,t)$ for finding the particle at time $t$ at coordinate
$x$, $\rhored$ being the reduced density matrix of the Brownian
particle.  Because the initial state $\rho_\mathrm{ini}$ describes a
superposition of the particle emanating from either slit, $P(x,t)$
displays a spatial interferece pattern, from which FLO extract an
attenuation factor $\aFLO(t)$.  The decay of $\aFLO(t)$ allows them to
define the decoherence time $\tdFLO$ discussed in the previous
section and below.

\section{\label{P_result} Derivation of $P(x,t)$}

Because the reader may not be familiar with the framework of the
quantum Langevin equation, we include a different derivation of $P(x,t)$ 
in a path integral framework in Appendix \ref{Grabert}.
The result is
\be
\label{Ptotal}
P(x,t) = 
\Pcl(x,t)
+ \Pint(x, t) \cos \left(\frac{xd A(t)}{2 \sigma^2 w(t)^2} \right), 
\ee
with
\bea
\label{Pcl}
\Pcl(x,t) 
&=&
\frac{N}{2}  \left( P_1(x - d/2, t) + P_1(x+d/2, t)
\right)  
\nonumber \\
&\equiv&
\frac{1}{2} \left( \Pcl^-(x, t) + \Pcl^+(x, t) \right),
\eea
\be
\label{P1}
P_1(x,t) = \frac{1}{\sqrt{2 \pi w(t)^2}} \cdot \exp \left(- \frac{x^2}{2 w(t)^2} \right),
\ee
\be
\label{Pint}
\Pint =  \frac{N}{\sqrt{2 \pi w(t)^2}} \cdot \exp \left(- \frac{x^2 + d^2 (\sigma^2 - 2 Q(t)) / (4
  \sigma^2)}{2 w(t)^2} \right), 
\ee 
the width $w(t)$ of the wave packets being
\be
\label{wsq}
w(t)^2 = \sigma^2 + \frac{A(t)^2}{\sigma^2} - 2 Q(t).
\ee
The quantities $A(t)$ and $Q(t)$ are defined as the imaginary
and real part of the position-position autocorrelation function
$\langle (x(t) - x(0)) x(0) \rangle \equiv Q(t) + i A(t)$, and
are related to the parameters in FLO's work by
\be
\label{FordGrabertRel}
[x(t_1), x(t_1+t)] = 2 i A(t), \;\;\; s(t) = - 2 Q(t).
\ee

As is shown in Appendix \ref{Grabert},
$P_1(x \mp d/2,t)$ is the probability distribution if only one slit,
centered around $x = \pm d/2$, was present;  
$\Pint(x,t)$ is the amplitude of the interference pattern.
The resulting expression (\ref{Ptotal}) for $P(x,t)$ agrees with FLO's result.

The explicit form of $A(t)$ and $Q(t)$ in the case of an Ohmic heat
bath with infinite cutoff and friction coefficient $\gamma$ is 
quite cumbersome and given in Eq. (9.14) and (9.15) of
\olcite{GrabertIngold88}.
However, for our purposes the limiting case  $\gamma \ll T$ will be
sufficient, which we assume from now on.
In this case, $A(t)$ and $Q(t)$ are given by
\be
A(t) =  \frac{1 - e^{- \gamma t}}{2 m \gamma}, \; \; 
Q(t) = -  \frac{T}{m \gamma} \left( t - \frac{1 - e^{- \gamma
    t}}{\gamma} \right).
\ee
For $\gamma = 0$, these equations reduce to
\be
A(t) = \frac{t}{2m}, \; \; Q(t) = - \frac{T t^2}{2m}
\ee

Above and hereafter, we chose units with $\hbar = k_B = 1$.
After the mass, length and energy scales are set by the particle mass
$m$, the distance of the slits $d$, and by $E \equiv m^{-1} d^{-2}$,
there are three remaining free parameters in the theory: the slit width
$\sigma / d$, the temperature $T / E$, and the friction coefficient of the
Ohmic heat bath $\gamma / E$. 
In all plots below, we set the slit width as $\sigma = d / 20$ unless
otherwise stated.

\section{\label{critique} Critique of FLO's analysis}

Before we comment on the further analysis of FLO, we shall briefly discuss
some properties of the probability density $P(x,t)$.
\begin{figure}
\epsfig{file=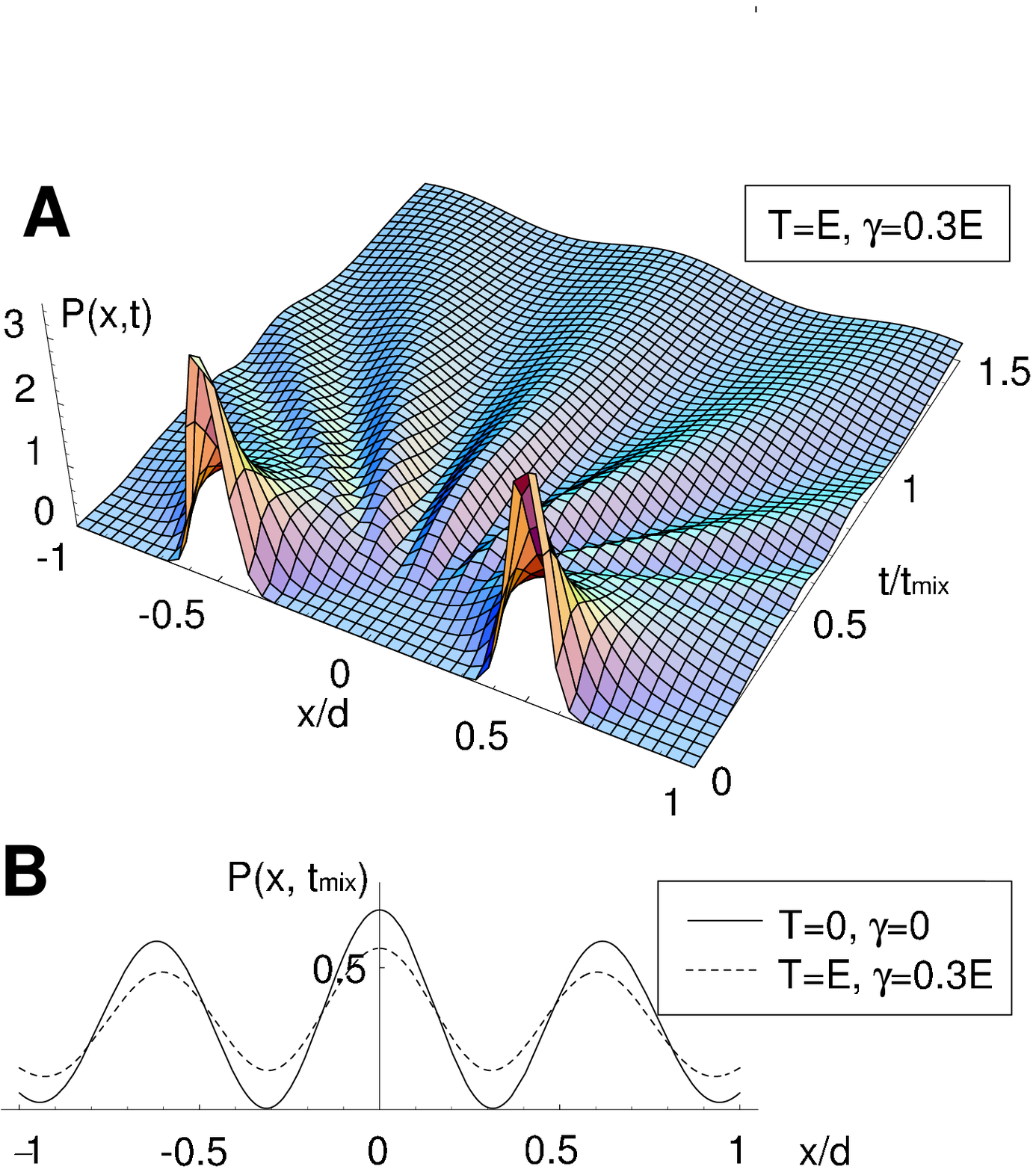, width=1\linewidth}
\caption{\label{Pxt_plot} 
A: The probability density  $P(x,t)$ for finding the particle at time $t$
at coordinate $x$ is plotted for $T = E, \gamma = 0.3 E$.
An interference fringe is seen to appear at time $\tmix$. 
B: $P(x,t = \tmix)$
is plotted for the same parameters as in A (dashed line), and for $T =
\gamma = 0 E$ (solid line).
The interference fringe in the former curve is seen
to be somewhat suppressed with respect to the latter, but to be
qualitatively very similar.
}
\end{figure}
In \Fig{Pxt_plot}A, $P(x,t)$ is plotted for the parameters
$\gamma = 0.3 E$, $T = E$ (in the following, these parameters will be
referred to as the weak dissipation case).
An interference pattern is seen to emerge only after the two wave
packets, initially separated by  $d$, have developed a  significant
overlap.
The associated time scale $\tmix$ is implicitly given by $w(\tmix) =
d$.
As long as friction and thermal spreading of the wave packets is
dominated by quantum broadening
($\gamma \ll T \ll E \cdot d^2 / \sigma^2$), $\tmix$ is
given by
\be
\label{tmix_def}
\tmix \equiv 2 m \sigma d,
\ee
which we will use as a definition of $\tmix$ from now on.
For $t < \tmix$, the interference pattern is influenced not only by
the loss of phase coherence, but also (and mainly) by the spreading of
the wave packets, as is shown below.
For $t > \tmix$, the interference pattern is seen to broaden and to
become flatter, as the wave function continues to spread.

\begin{figure}
\epsfig{file=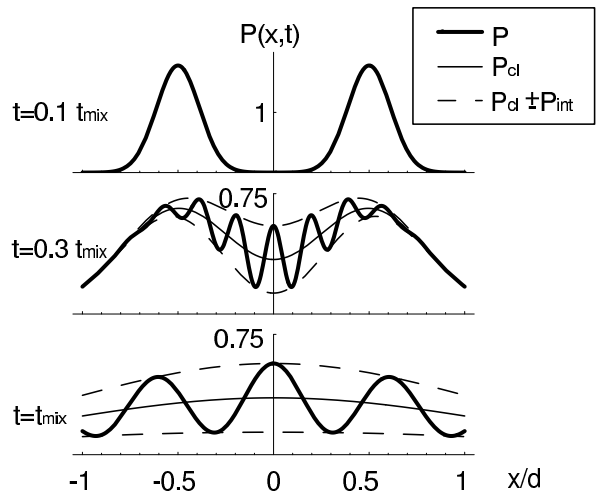, width=.9\linewidth} 
\caption{\label{Px_tshort} 
$P(x,t)$ is shown at the times $t = 0.1 \cdot \tmix$;  $t = 0.3
\cdot \tmix$; and  $t = \tmix$ (from top to bottom), for  
the parameters $T = E$, $\gamma = 0.3 E$ (thick line).
Also shown: The noninterfering contribution $\Pcl(x,t)$ (thin
line) and the envelope $\Pcl(x,t) \pm \Pint(x,t)$ 
of the interference pattern (dashed line) around $\Pcl(x,t)$.
}
\end{figure}

In \Fig{Pxt_plot}B, the interference pattern 
at time $t=\tmix$ in the weak dissipation case is compared to the case 
$T = \gamma = 0$.
In \Fig{Px_tshort}, $P(x,t)$ is shown
for three different times,
again using the parameters of the weak dissipation case, together with
the noninterfering part of the
amplitude $\Pcl(x, t)$ and with the 
envelope of the interference pattern, given by 
$\Pcl (x,t) \pm \Pint (x,t)$.
As can be seen in \Fig{Px_tshort}, $\Pcl(x=0,t)$ is vanishingly small
at $t \ll \tmix$, and is rapidly growing as the wave packets start to overlap.
When the temperature is further increased, $\Pcl(x=0,t)$ grows even faster for
$t < \tmix$ due to the additional thermal spreading of the wave
packets (not shown).


In \Fig{Pxt_plot}B, the interference fringes for the case of
no dissipation on the one hand and for weak dissipation on the other
hand look quite similar.
This is in drastic contrast to what one might expect from FLO's
analysis, which leads to a decoherence time given by
\be
\tdFLO = \frac{s^2 m^{1/2}}{d T^{1/2}}.
\ee
The parameters chosen for the weak dissipation case in \Fig{Pxt_plot}B
imply, for example, $\tdFLO = 0.025 \cdot \tmix$ (extracted from
the very same function $P(x,t)$!). 
We believe this value implies that by the time $\tmix$, the  
entire interference  
pattern, clearly visible in \Fig{Pxt_plot}, should have already disappeared. 
How can this be? 
 
\begin{figure}
\epsfig{file=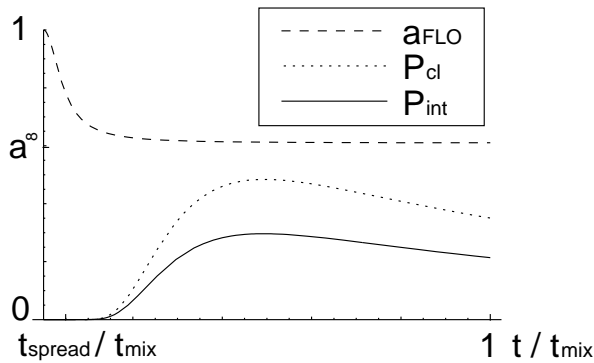, width=.9\linewidth} 
\caption{\label{PPa_t} 
$\Pcl(x=0,t)$ and $\Pint(x=0,t)$ (dotted and solid line)
are shown along with their ratio $\aFLO = \Pint / \Pcl$
(dashed line) for $T=E$, $\gamma = 0$. 
Both $\Pint$ and $\Pcl$  grow rapidly after time
$t_\mathrm{spread}$, after which thermal broadening of the wave
packets begins.  
$\aFLO$ is seen to drop only for $t < t_\mathrm{spread}$.
}
\end{figure}

The decoherence analysis of FLO is based on an
attenuation factor $a_\mathrm{FLO}(t)$, which is defined as ``the
ratio [...] of the amplitude of the interference term to twice the
geometric mean of the other two terms'' \cite{FordOConnell01}, i.e.~as  
\begin{subequations}
\label{eq:aflo}
\be 
a_\mathrm{FLO}(t) = \frac{\Pint(x,t)}{\sqrt{\Pcl^+(x,t) \cdot
    \Pcl^-(x,t)}}.  
\ee 
An example of $\aFLO(t)$ is shown
in  Fig.~\ref{PPa_t} (dashed line). 
Using Eqs.~(\ref{Pcl}) to (\ref{Pint}),
$a_\mathrm{FLO}(t)$ can be recast in the form 
\be
\label{a_FLO}
a_\mathrm{FLO}(t) = \frac{\Pint(x=0,t)}{\Pcl(x = 0, t)}.  
\ee 
\end{subequations}
In other words, $\aFLO$ measures the interference amplitude $\Pint$ in
units of the classical amplitude $\Pcl$ at $x=0$;
hence it does not merely measure the time dependence
of the interference pattern, but also reflects the drastic increase of
the reference unit $\Pcl(x=0,t)$ for $t < \tmix$.

This is illustrated in \Fig{PPa_t}, where $\Pint(x=0,t)$ and
$\Pcl(x=0,t)$ are plotted along with 
their ratio $a_\mathrm{FLO}(t)$ for finite temperature $T=E$ and no
dissipation ($\gamma =0$).
Using \Eq{P1}, (\ref{Pint}), the exponents in $\Pint$ and $\Pcl$
are seen to be only slowly departing from their initial values
$\Pint(t=0), \Pcl(t=0) \sim \exp(-d^2/8\sigma^2)$ ($\sim 10^{-22}$ for the given
parameters!) as long as $t < t_\mathrm{spread} \equiv \tmix \; \sigma / d$.
This is because for $t < t_\mathrm{spread}$ the wave packet width $w(t)$ 
in \Eq{wsq} is dominated by the constant $\sigma^2$, and that quantum
spreading is not effective yet.
The rapid growth of $\Pint(t)$ and $\Pcl(t)$ by 22 orders of magnitude
seen in \Fig{PPa_t} takes place almost entirely between
$t_\mathrm{spread}$ and $\tmix$, when the overlap of the wave packets
increases rapidly  due to quantum spreading.
On the other hand, the decrease of $\aFLO$ takes place before
$t_\mathrm{spread}$, when  $\Pint$ and $\Pcl$ are still tiny.
Indeed,  $t < t_\mathrm{spread}$ is precisely the
condition that  $\aFLO$ can be fitted by a Gaussian, $\aFLO \approx
\exp(-t^2 / 8 (\tdFLO)^2)$, from which $\tdFLO$ was extracted by FLO.
Once quantum broadening sets in after $t_\mathrm{spread}$,  $\aFLO$
crosses over to a constant.

In summary, $\aFLO$ is seen to reflect mainly the details of the
broadening of the wave packets, and therefore does not appear to us to
be a suitable measure for their coherence.  

\section{\label{alternative} Alternative analysis of $P(x,t)$}

As long as the overlap of the two wave packets is held fixed, the
amplitude of the interference pattern 
$\Pint(x,t)$  is, indeed, a direct measure of their phase coherence.
Similar interference patterns have been analyzed to this purpose in a
number of very illuminating experiments, most explicitly in
\olcite{KokorowskiPritchard01}.
In the thought experiment considered here,
the wave packets overlap only after a time $\tmix$.
This introduces a considerable amount of arbitrariness in a
definition of decoherence for times 
$t < \tmix$, if this definition is based upon the diagonal elements of the
reduced density matrix $P(x,t) =  \tilde{\rho}(x,x,t)$ only.
For the regime $t < \tmix$, an unambiguous measure of decoherence can
only be obtained from the decay of the off-diagonal elements of the
reduced density matrix $\tilde{\rho}$, as for example in
\olcite{StrunzHaake03}. 
However, this would go beyond the scope of the present article, and
such an analysis is only done for the dissipationless case $\gamma
=0$ (in section \ref{OD}).
In this section, we take a less ambitious approach and just
give an example showing how, for
these short times, a different definition of an
attenuation factor based on the probability density (i.e.~the diagonal
elements of $\tilde{\rho}$)
results in a picture completely different from that of FLO.

For this purpose, we introduce the attenuation factor
\be
\label{a2_def}
a_2(t) = \frac{\Pint (x=0,t) }{ N \cdot P_1 (x=0, t) },
\ee
where  $P_1$ and $\Pint$ are defined in \Eq{P1} and \Eq{Pint}, and $N$
is the trivial normalization factor defined after \Eq{alpha_2slit}.
Similarly to $\aFLO$,
this measures the relative importance of the interference amplitude $\Pint$
with respect to the noninterfering contribution.
The difference with respect to $\aFLO(t)$ in \Eq{a_FLO} is that
the relative importance of the noninterfering part of $P(x,t)$ 
is captured in a slightly different (but by no means less arbitrary) way.
In order to make the connection to the quantities shown in
\Fig{Px_tshort},
$N \cdot P_1(x=0,t)$ is proportional to the heigth of
$\Pcl$ at the center of one of the slits  (i.e.~ at $x = \pm 0.5 d$),
whereas in the definition of $\aFLO$ the value at $x = 0$ was taken as
reference unit.

We would like to emphasize that our $a_2$ is not ``better'' or ``more
appropriate'' than $\aFLO$. 
Indeed, for $t > \tmix$, $a_2$ and $\aFLO$, and
hence all conclusions drawn from them, are the same.
For short times $t < \tmix$, however,   $\aFLO$ and $a_2$ differ
wildly; this illustrates that  on these
short time scales, {\emph{both}} definitions are dominiated by
effects of wave packet spreading rather than decoherence, albeit in a different way.
We would like to emphasize that the condition $\aFLO(t=0) = 1$ that
distinguishes $\aFLO$ from $a_2$ is in our opinion not necessary,
because at times $t\ll \tmix$, the attenuation factors have nothing to
say about decoherence anyway.

\begin{figure}
\epsfig{file=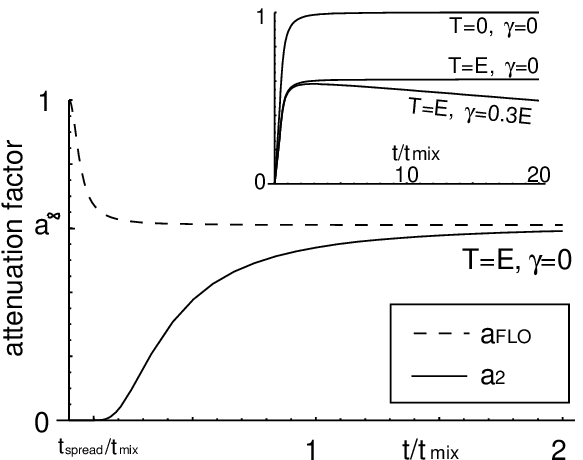, width=.9\linewidth} 
\caption{\label{a2_t} 
The attenuation factors $a_2(t)$ from \Eq{a2_def} (solid line) and
$\aFLO(t)$ from \Eq{a_FLO} (dashed line) are compared and seen to be
wildly different for $t < \tmix$.
$a_2$ is related to $\Pint$ in \Fig{PPa_t}, the only difference being
that the prefactor $(2 \pi w^2)^{-1/2}$ present in \Eq{Pint} for
$\Pint$ is missing in $a_2$. 
Inset: The attenuation factor $a_2(t)$ is plotted as a function of time, for $T = \gamma = 0$, for
  $T=E$, $\gamma = 0$, and for $T = E$, $\gamma = 0.3 E$ (from top to bottom).
}
\end{figure} 
In \Fig{a2_t}, the time evolution of
$a_2$ and $\aFLO$ are compared for finite temperature $T=E$ and
$\gamma =0$.
$\aFLO$ is seen to decay from the initial value
$\aFLO(t=0) = 1$, whereas $a_2$ is growing,
more closely resembling the actual interference pattern seen in
\Fig{Pxt_plot}.
For $t > \tmix$, both $\aFLO$ and $a_2$ become equal and
saturate at the value
\be
\label{a2sat_T}
a^\infty = \exp \left(- \frac{d^2}{8 \sigma^2 + 2 \lth^2} \right),
\ee
where $\lth^2 = 1 /(m T)$ is the squared thermal wavelength.
As is shown in section \ref{OD} below, the reduction of
$a^\infty$ at finite temperature 
results from the imperfect preparation of the initial state, and has
no time scale associated with it, contrary to what the 
time evolution of $\aFLO$ might suggest.

Such a time scale is only introduced when $\gamma > 0$:
As is shown in the inset of \Fig{a2_t}, in this case $a_2$ and
$\aFLO$ is further reduced in a time-dependent way and assumes for
$t \gg \mathrm{max}( 1/\gamma, \tmix)$
the simple limiting form
\be 
\label{a2_exp}
\aFLO(t), a_2(t) \rightarrow \exp \left( - \frac{t}{\tdec ( 1 +
  t/t_s)} \right),
\ee
where
$\tdec = \lth^2/ (d^2 \gamma ) $, and 
$t_s = \tdec  \, d^2 / (8 \sigma^2) \gg \tdec$.

Remarkably, for $t \gg t_s$,  $a_2$ and $\aFLO$ are found to saturate at the (tiny) value 
$a_2(t \rightarrow \infty ) = \exp( - d^2 / (8\sigma^2))$.
This is probably related to the small initial overlap of the wave packets,
i.e.~to the part
of the wave function, for which the which-path information cannot be
distinguished by the environment.
An analogous saturation was found in the case of
a particle in a harmonic potential.\cite{CaldeiraLeggett85}

For $\tmix < t \ll t_s$, $a_2$ and $\aFLO$ decay exponentially on the
decoherence time 
scale $\tdec$ (this was also found by FLO in the long time limit).
The decoherence time $\tdec$ agrees with what one expects on
general grounds:
The paths emanating from either slit aquire random phase
differences, which depend linearly on the bath coupling $\gamma$ and
rise quadratically with their distance $d$.
This functional dependence of $\tdec$ has both been predicted
theoretically \cite{Zurek91} and  observed experimentally in
a somewhat similar context.\cite{KokorowskiPritchard01}
Note that $\tdec$ diverges as
$\gamma$ or $T$ vanish, such that no decoherence without dissipation
is seen.

\section{\label{OD}Short-time analysis of the dissipationless case}

We have argued that neither $a_2$ nor $\aFLO$ are suitable to reveal
meaningful information about decoherence at $t < \tmix$.
How, then, would one obtain such information?
Because this question addresses time scales shorter than $\tmix$, it
cannot be answered using the diagonal elements of the reduced
density matrix $\tilde{\rho}$ alone,
on which the attenuation factors  $\aFLO$ and $a_2$ are based.
Instead, it is instructive to look at the entire reduced density
matrix, including its off-diagonal elements.
A definition of decoherence valid for all times including $t < \tmix$
has been proposed in Ref.~\olcite{StrunzHaake03}.
It relies on the observation that $\rhored$ splits naturally into a
classical part $\rhored_\mathrm{cl}$ and an interference part
$\rhored_\mathrm{int}$ 
($\rhored = \rhored_\mathrm{cl} +\rhored_\mathrm{int}$), 
which produce the corresponding 
terms in the probability density in \Eq{Ptotal} (this is seen
explicitly in \Eq{alphaalpha_2slit}).
Therefore, the norm $a_\mathrm{OD}(t)$, defined in
Ref.~\olcite{StrunzHaake03} by
\be
\label{aOD_def}
|a_\mathrm{OD}(t)|^2 = 
\mathrm{Tr} \rhored_\mathrm{int}(t)^{\phantom{\dagger}} \!\!
            \rhored_\mathrm{int}(t)^\dagger,
\ee
describes the temporal fate of the interference term even for
$t < \tmix$, i.e.~before it appears in the probability density.
Very importantly,
the dissipationless case $\gamma = 0$ describes a closed system with
unitary time evolution, $\rhored(t) = 
U^{\phantom{\dagger}}\!\! \rhored U^\dagger$ with $U^{-1} = U^\dagger$;
hence \Eq{aOD_def} is in this case automatically independent of time!
This is the back-of-the-envelope ``proof'' (already given in
Ref.~\olcite{StrunzHaake03}) that there can be no decoherence without
dissipation.

Further insight is gained by calculating the value of
$a_\mathrm{OD}$ for $\gamma = 0$, which is conveniently done at time
$t=0$.
The initial wave function for fixed momentum $p$ is given by
\be
\psi_p(x) = \sqrt{\frac{N}{2}} e^{i p x}
\left( \psi_+(x) + \psi_-(x) \right).
\ee
where $N$ is given after \Eq{alpha_2slit}, and where
\be
\psi_\pm(x) = \frac{1}{(2 \pi \sigma^2)^{1/4}} 
e^{ - \frac{(x \pm d/2)^2}{4 \sigma^2} }.
\ee
Correspondingly, the initial density matrix at temperature $T$ is
given by \cite{FordOConnell01c}
\begin{eqnarray}
\label{rho_calc}
\rho(x,x') &=& \frac{N \lth}{\sqrt{8 \pi}} \int dp
e^{ - p^2 \lth^2 / 2  + i p x }
\sum_{i,j = \pm} \psi_i(x) \psi_j^*(x') \nonumber \\
&=& \frac{N}{2} e^{ - \frac{(x - x')^2}{2 \lth^2} }
\sum_{i,j = \pm} \psi_i(x) \psi_j^*(x'),
\end{eqnarray}
$e^{- p^2 \lth^2 / 2 }$ being the Boltzmann factor.
The interfering part $\rhored_\mathrm{int}$ of $\rhored$ is that part
of \Eq{rho_calc} for which the indices $i$ and $j$ represent
different signs ($i = -j$).
In the limit $\sigma \ll \lth, d$, it follows that
\be
\label{a_proof}
a_\mathrm{OD}(t) = \frac{1}{\sqrt{2}} \exp\left(- \frac{d^2}{2 \lth^2}
\right) =  \frac{1}{\sqrt{2}} a^\infty.
\ee
(Only the temperature dependence of $a_\mathrm{OD}$ is important; the
trivial factor $1/\sqrt{2}$ would disappear if $a_\mathrm{OD}$ was
normalized differently, e.g.~by dividing \Eq{aOD_def} by
the corresponding ``classical'' quantity
$\mathrm{Tr} \rhored_\mathrm{cl}(t)^{\phantom{\dagger}} \!\!
\rhored_\mathrm{cl}(t)^\dagger$).
As was seen on general grounds, this is independent of time.
Consequently,  in the absence of dissipation, {\emph{no time scale}} is
associated with the reduction of the attenuation factor  $a^\infty$
below 1 as the temperature is increased.

Instead, this reduction is already present in the initial state,
and there is a very simple explanation for it that has nothing to do
with decoherence: 
In \Eq{rho_calc}, the initial density matrix is seen to have four
peaks, two diagonal ones around $x = x' = \pm d/2$ (belonging to
$\rho_\mathrm{cl}$), and two off-diagonal ones at $x = -x' = \pm d/2$
(belonging to $\rho_\mathrm{int}$; these latter peaks give rise to the
interference pattern after they have 
spread enough to show up in the diagonal elements of $\rhored$, which is
another way of understanding the origin of the mixing time
$\tmix$).
It is seen in \Eq{rho_calc} that as a consequence of the Boltzmann
factor, the off-diagonal peaks in \Eq{rho_calc} are suppressed 
with respect to the diagonal ones by precisely the factor
$a^\infty$ as given in \Eq{a_proof}.
Thus, the reduction of the interference pattern
at nonzero temperature simply results from
the fact that the initial state is not a pure
state, but an  imperfectly prepared mixed state
with a momentum uncertainty of the order of $\lth^{-1}$.
Of course, the imperfect preparation of the initial state should not
be confused with decoherence: The latter is a dynamical process with
an associated time scale, the former is not.


\section{Conclusion}
The main result of this work is contained in the figures which clearly show
coherence on time scales greater than $\tdFLO$.  It is important to
emphasize that the formulas used to obtain these figures are identical
to FLO's, rederived here for completeness by another method. 
For short times $t < \tmix$,
the measure of decoherence suggested by FLO
does not permit 
the  separation of the change in overlap of the wave packets from the decay of the 
interference pattern, and therefore has nothing to say about
decoherence.
However, a simple calculation of the attenuation factor based on the
off-diagonal elements of the reduced density matrix at $\gamma = 0$
clearly shows that it does not depend on time at all.
Decoherence without dissipation? We think not.


\begin{acknowledgments}
The work of DG and VA has been supported in part by the NSF grant DMR-0242120.
DG acknowledges partial travel funding from CeNS at LMU M\"unchen.
We thank F.~Wilhelm, M.~Thorwart, G.~Ford and R.~O'Connell for
discussions.
The figures were created using Mathematica.
\end{acknowledgments}

{\bf Note added:} From a private communication with Ford, we learnt that
Murakami, Ford and O'Connell \cite{MurakamiOConnell03} 
have recently concluded themselves that
in the absence of dissipation, ``there is no decoherence in (Wigner) phase
space''. We fully agree with this conclusion, but (contrary to
\olcite{MurakamiOConnell03}) believe 
it to be inconsistent with FOL's earlier claims of decoherence
without dissipation (in coordinate space). The resolution
of this inconsistency is that FOL's measure of decoherence
in coordinate space is meaningless in the short-time limit,
as argued in the main text.

\appendix

\section{\label{Grabert} Derivation of $P(x,t)$ using path integrals}

A free Brownian particle, coupled to a bath of harmonic oscillators, can
be solved exactly by a number of methods, such as path integral
\cite{GrabertIngold88} or operator methods \cite{HakimAmbegaokar85}.
We will give a brief account of an alternative derivation of
the probability density $P(x,t) = \rhored(x,x,t)$ using a path
integral approach. 
The entire appendix relies heavily on the concepts and results from
previous work by Grabert an collaborators \cite{GrabertIngold88}, to
which will be referred frequently.

In the path integral framework,
the reduced density matrix at time $t_f$ is given by
\begin{widetext}
\bea
\label{full_PI}
\rhored(x_f, x_f', t_f) =  
\prod_\alpha  \int dQ_{\alpha, f} \cdot 
\Dpath{x}{\mathrm{any}}{x_f} \Dpath{x'}{\mathrm{any}}{x_f'}
\Dpath{Q_\alpha}{\mathrm{any}}{Q_{\alpha, f}}
\Dpath{Q_\alpha'}{\mathrm{any}}{Q_{\alpha, f}}  \cdot
\nonumber \\
e^{i S [x(\cdot), \{ Q_\alpha(\cdot) \} ]} e^{- i S [x'(\cdot), \{ Q_\alpha'(\cdot) \} ]}
\cdot \alpha^*\left(x(0)\right) \alpha\left(x'(0)\right) \cdot 
\rhoth \left(x(0),x'(0),  \{ Q_\alpha(0) \}, \{ Q_\alpha'(0) \}
\right).
\eea
Several symbols used here require some explanation.
$Q_\alpha$ and $Q_\alpha'$ are the coordinates of the 
harmonic oscillators representing the environment
degrees of freedom, labelled by the index $\alpha$.
$\Dpath{x}{x_i}{x_f}$ is an integration over all paths, i.e.~functions
of time $0 \leq t \leq t_f$, with 
boundary values $x(0) = x_i$,  $x(t_f) = x_f$; 
the integration boundary ``any'' indicates that
an integration over all coordinate values at the boundary is
performed.
The integration over $Q_{\alpha, f}$
performs the trace over the environment degrees of freedom.
Details about path integrals can be found in \olcite{FeynmanHibbs65}.

Following \olcite{CaldeiraLeggett83},
the action $S [x, \{ Q_\alpha \} ]$ for system and
environment is given by 
\be
\label{action}
S  [x, \{ Q_\alpha \} ] = 
\int_0^{t_f} dt \frac{m}{2} \dot{x}(t)^2+ 
\sum_\alpha \left( 
\frac{m}{2} \dot{Q}_\alpha(t)^2 - \frac{m_\alpha \omega_\alpha^2}{2} 
\left( Q_\alpha(t) - \frac{C_\alpha }{m_\alpha \omega_\alpha^2} x(t) \right)^2
\right).
\ee
The mass $m_\alpha$, coupling $C_\alpha$ and frequency $\omega_\alpha$
of the environment oscillators enter the reduced density matrix
only via the spectral function 
$J(\omega) \equiv \frac{\pi}{2} \sum_\alpha \frac{C_\alpha^2}{m_\alpha
  \omega_\alpha} \delta(\omega - \omega_\alpha)$,
which we take to have the appropriate form for Ohmic damping, 
$J(\omega) = m \gamma \omega$.
Strictly speaking, $J(\omega)$ has to be cut off at high frequencies.
In the quantities we are interested in, however, no divergencies 
are encountered as this cutoff is taken to infinity.

The thermal density matrix $\rhoth$ can be evaluated using an
imaginary-time path integral,
\be
\label{rhoth}
\rhoth \left(x,x',  \{ Q_\alpha \}, \{ Q_\alpha' \} \right) =  
\frac{1}{Z}
\Dpath{\xbar}{x}{x'}
\prod_\alpha
\Dpath{\Qbar_\alpha}{Q_\alpha}{Q_\alpha'}
e^{- S^E [\xbar(\cdot), \{ \Qbar_\alpha(\cdot) \} ]},
\ee
where the path integral is performed over paths with imaginary-time
argument  $i \tau \in [0 , i/T]$,
$-S^E$, explicitly
given by Eq.~(3.3) of \olcite{GrabertIngold88}, is the action $iS$ (\ref{action}),
analytically continued to the imaginary time $i \tau$, and
$Z$ is the partition function, such that $\mathrm{tr}\rhoth = 1$.

From \Eq{full_PI} - (\ref{rhoth}), a propagation function $J$ for the
reduced density matrix $\rhored$ can be defined,
\begin{eqnarray}
\label{J_def}
J(x_f, x_f', t_f, x_i, x_i') =  
\frac{1}{Z}
\prod_\alpha  \int Q_{\alpha, f} \cdot
\Dpath{x}{\mathrm{any}}{x_f} \Dpath{x'}{\mathrm{any}}{x_f'}
\Dpath{Q_\alpha}{\mathrm{any}}{Q_{\alpha, f}}
\Dpath{Q_\alpha'}{\mathrm{any}}{Q_{\alpha, f}}  \nonumber \\
\Dpath{\xbar}{x(0)}{x'(0)}
\Dpath{\Qbar_\alpha}{Q_\alpha(0)}{Q_\alpha'(0)} 
\cdot
e^{i S [x, \{ Q_\alpha \} ]}
e^{- i S [x', \{ Q_\alpha' \} ]}
e^{- S^E [\xbar, \{ \Qbar_\alpha \} ]},
\end{eqnarray}
such that
\be
\label{rho_timeevo}
\rhored(x_f, x_f', t_f) =  
\int dx_i dx_i'
J(x_f, x_f', t_f, x_i, x_i')
\cdot \alpha^*\left(x_i\right) \alpha\left(x_i'\right).
\ee
\end{widetext}

A similar propagation function $J^F$ is also defined in Eq.~(3.35) of
\olcite{GrabertIngold88}.
There, however, a  somewhat more general class of initial state
preparations was considered, resulting in two more arguments $\xbar$
and $\xbar'$ of $J_F$.
The propagation function $J$ in \Eq{J_def} is related to $J^F$ by  
$J(x_f, x_f', t_f, x_i, x_i') = 
J^F(x_f, x_f', t_f, x_i, x_i', \xbar_i = x_i,  \xbar_i' =x_i')$.

In the case of a free particle coupled to an Ohmic heat bath,
the propagating function $J$ in \Eq{J_def} can be explicitly
evaluated. 
In fact, this has kindly been done in \olcite{GrabertIngold88}, and the
result is given by Eq. (9.14) and (9.15) there.
For the diagonal element
$\rhored(x_f, x_f, t_f)$ of the reduced density matrix, only the
elements of $J$ with $x_f' = x_f$ are needed , which are given by
\bea
\label{J0}
J_0(X,Y, t_f)
&\equiv&
J(x_f, x_f, t_f, x_i, x_i') 
\nonumber \\
&=& \frac{1}{4\pi A(t)} 
e^{i \frac{X Y}{2 A(t_f)} +   X^2 \frac{Q(t_f)}{4 A(t_f)^2}}.
\eea
Here, $X = x_i - x_i'$, $Y = x_f - (x_i + x_i')/2$, and
$A(t)$ and $Q(t)$ are the imaginary
and real parts of the position-position autocorrelation function
$\langle (x(t) - x(0)) x(0) \rangle = Q(t) + i A(t)$.
$A(t)$ and $Q(t)$ are given by Eq. (10.1) and (10.4) of \olcite{GrabertIngold88},
and related to the parameters in FLO's work by \Eq{FordGrabertRel}.
\Eq{rho_timeevo} and \Eq{J0}, with $Y - x_f$ substituted by $q$, can be
recast in the compact form 
\bea
\label{P_solution}
P(x_f,t) &=& \int d X d q J_0(X_i, x_f + q, t) 
\nonumber \\
&\cdot& \alpha^*(q - X / 2) \alpha(q + X / 2).
\eea

\Eq{P_solution} allows to rederive
the results obtained by FLO. First, let us consider the time evolution
of a wave function emerging from a single Gaussian slit, described by
the preparation function
\be
\alpha(x) = \frac{1}{ (2 \pi \sigma^2)^{1/4} } 
e^{- \frac{x^2}{4 \sigma^2} },
\ee
such that
\begin{widetext}
\be
\label{alphaalpha_1slit}
\alpha^*(q - X / 2) \alpha(q + X / 2) 
=
\frac{1}{ \sqrt{2 \pi \sigma^2} } 
e^{ - \frac{q^2}{2\sigma^2} } e^{ - \frac{X^2}{8\sigma^2} }
\equiv
t_1(X,q).
\ee
\Eq{P_solution}, being a Gaussian integral, can now easily be
evaluated and yields the result $P_1(x,t)$ given in \Eq{P1}.

In the case of a two-slit preparation with $\alpha(x)$ given by
\Eq{alpha_2slit},
\be
\label{alphaalpha_2slit}
\alpha^*(q - X / 2) \alpha(q + X / 2) 
=
\frac{N}{2} 
\left( t_1(X,q - \frac{d}{2}) + t_1(X,q + \frac{d}{2}) 
+ t_1(X - d, q) +  t_1(X + d, q) \right)
\ee
\end{widetext}
with $t_1$ defined in \Eq{alphaalpha_1slit}.
This is a sum of four terms, the 
first two of which  describe the emergence of both the forward and
the backward time evolution of the density matrix from the same slit,
whereas the remaining two terms describe the emergence from two
different slits.
The former two terms describe the ``classical'' sum of the
probabilities from each slit, whereas the latter ones are responsible
for the interference pattern.
By linearity of \Eq{rho_timeevo}, also $P(x,t)$ can be decomposed into a sum of four
corresponding terms, as in \Eq{Ptotal}.

The first two of these terms can be evaluated as in the case of a single slit,
with $q$ replaced by $q \pm d/2$. 
Their contribution to $P(x,t)$ is thus given by $\Pcl(x,t)$ as defined
in \Eq{Pcl}.

The remaining two terms of \Eq{alphaalpha_2slit} are mutually related by complex
conjugation.
The corresponding terms of $\Pcl(x,t)$ are easily evaluated using
\Eq{P_solution}; the result is
$\Pint(x,t)  \cdot  \cos \left( \frac{ A(t) \cdot d \cdot x}{2 w(t)^2 \cdot \sigma^2} \right)$,
with $\Pint$ given in \Eq{Pint}.
The resulting expression for $P(x,t)$ is given in \Eq{Ptotal}. 
Using \Eq{FordGrabertRel}, it is seen to agree with FLO's result.


\begin{thebibliography}{17}
\expandafter\ifx\csname natexlab\endcsname\relax\def\natexlab#1{#1}\fi
\expandafter\ifx\csname bibnamefont\endcsname\relax
  \def\bibnamefont#1{#1}\fi
\expandafter\ifx\csname bibfnamefont\endcsname\relax
  \def\bibfnamefont#1{#1}\fi
\expandafter\ifx\csname citenamefont\endcsname\relax
  \def\citenamefont#1{#1}\fi
\expandafter\ifx\csname url\endcsname\relax
  \def\url#1{\texttt{#1}}\fi
\expandafter\ifx\csname urlprefix\endcsname\relax\def\urlprefix{URL }\fi
\providecommand{\bibinfo}[2]{#2}
\providecommand{\eprint}[2][]{\url{#2}}

\bibitem[{\citenamefont{Ford et~al.}(2001)\citenamefont{Ford, Lewis, and
  O'Connell}}]{FordOConnell01}
\bibinfo{author}{\bibfnamefont{G.}~\bibnamefont{Ford}},
  \bibinfo{author}{\bibfnamefont{J.}~\bibnamefont{Lewis}}, \bibnamefont{and}
  \bibinfo{author}{\bibfnamefont{R.}~\bibnamefont{O'Connell}},
  \bibinfo{journal}{Phys. Rev. A} \textbf{\bibinfo{volume}{64}},
  \bibinfo{pages}{032101} (\bibinfo{year}{2001}).

\bibitem[{\citenamefont{see~e.g. U.~Weiss}(1993)}]{Weiss93}
\bibinfo{author}{\bibnamefont{see~e.g. U.~Weiss}},
  \emph{\bibinfo{title}{Quantum Dissipative Systems}}
  (\bibinfo{publisher}{World Scientific}, \bibinfo{year}{1993}).

\bibitem[{\citenamefont{Chiorescu et~al.}(2003)\citenamefont{Chiorescu,
  Nakamura, Marmans, and Mooij}}]{ChiorescuMooij03}
\bibinfo{author}{\bibfnamefont{I.}~\bibnamefont{Chiorescu}},
  \bibinfo{author}{\bibfnamefont{Y.}~\bibnamefont{Nakamura}},
  \bibinfo{author}{\bibfnamefont{C.}~\bibnamefont{Harmans}}, \bibnamefont{and}
  \bibinfo{author}{\bibfnamefont{J.}~\bibnamefont{Mooij}},
  \bibinfo{journal}{Science} \textbf{\bibinfo{volume}{299}},
  \bibinfo{pages}{1869} (\bibinfo{year}{2003}).

\bibitem[{\citenamefont{Nakamura et~al.}(1999)\citenamefont{Nakamura, Pashkin,
  and Tsai}}]{NakamuraTsai99}
\bibinfo{author}{\bibfnamefont{Y.}~\bibnamefont{Nakamura}},
  \bibinfo{author}{\bibfnamefont{Y.}~\bibnamefont{Pashkin}}, \bibnamefont{and}
  \bibinfo{author}{\bibfnamefont{J.}~\bibnamefont{Tsai}},
  \bibinfo{journal}{Nature} \textbf{\bibinfo{volume}{398}},
  \bibinfo{pages}{786} (\bibinfo{year}{1999}).


\bibitem[{\citenamefont{Kokorowski et~al.}(2001)\citenamefont{Kokorowski,
  Cronin, Toberts, and Pritchard}}]{KokorowskiPritchard01}
\bibinfo{author}{\bibfnamefont{D.}~\bibnamefont{Kokorowski}},
  \bibinfo{author}{\bibfnamefont{A.}~\bibnamefont{Cronin}},
  \bibinfo{author}{\bibfnamefont{T.}~\bibnamefont{Toberts}}, \bibnamefont{and}
  \bibinfo{author}{\bibfnamefont{D.}~\bibnamefont{Pritchard}},
  \bibinfo{journal}{Phys. Rev. Lett.} \textbf{\bibinfo{volume}{86}}, \bibinfo{pages}{2191}
  (\bibinfo{year}{2001}).

\bibitem[{\citenamefont{Arndt et~al.}(1999)\citenamefont{Arndt, Nairz,
  Vos-Andreae, Keller, van~der Zouw, and Zeilinger}}]{ArndtZeilinger99}
\bibinfo{author}{\bibfnamefont{M.}~\bibnamefont{Arndt}},
  \bibinfo{author}{\bibfnamefont{O.}~\bibnamefont{Nairz}},
  \bibinfo{author}{\bibfnamefont{J.}~\bibnamefont{Vos-Andreae}},
  \bibinfo{author}{\bibfnamefont{C.}~\bibnamefont{Keller}},
  \bibinfo{author}{\bibfnamefont{G.}~\bibnamefont{van~der Zouw}},
  \bibnamefont{and}
  \bibinfo{author}{\bibfnamefont{A.}~\bibnamefont{Zeilinger}},
  \bibinfo{journal}{Nature} \textbf{\bibinfo{volume}{401}},
  \bibinfo{pages}{680} (\bibinfo{year}{1999}).

\bibitem[{\citenamefont{see e.g.~A.~Venugopalan}(1999)}]{Venugopalan99}
\bibinfo{author}{\bibnamefont{see e.g.~A.~Venugopalan}},
  \bibinfo{journal}{Phys. Rev. A} \textbf{\bibinfo{volume}{61}},
  \bibinfo{pages}{012102} (\bibinfo{year}{1999}).

\bibitem[{\citenamefont{Ford and O'Connell}(2001)}]{FordOConnell01b}
\bibinfo{author}{\bibfnamefont{G.}~\bibnamefont{Ford}} \bibnamefont{and}
  \bibinfo{author}{\bibfnamefont{R.}~\bibnamefont{O'Connell}},
  \bibinfo{journal}{Physics Letters A} \textbf{\bibinfo{volume}{286}},
  \bibinfo{pages}{87} (\bibinfo{year}{2001}).

\bibitem[{\citenamefont{Amvegaokar}(1993)}]{ambeg93}
\bibinfo{author}{\bibfnamefont{V.}~\bibnamefont{Ambegaokar}},
  \bibinfo{journal}{Phys.Today} \textbf{\bibinfo{volume}{46(4)}},
  \bibinfo{pages}{82} (\bibinfo{year}{1993}).

\bibitem[{\citenamefont{Caldeira and Leggett}(1985)}]{CaldeiraLeggett85}
\bibinfo{author}{\bibfnamefont{A.}~\bibnamefont{Caldeira}} \bibnamefont{and}
  \bibinfo{author}{\bibfnamefont{A.}~\bibnamefont{Leggett}},
  \bibinfo{journal}{Phys. Rev. A} \textbf{\bibinfo{volume}{31}},
  \bibinfo{pages}{1059} (\bibinfo{year}{1985}).

\bibitem[{\citenamefont{Ford and Kac}(1987)}]{FordKac87}
\bibinfo{author}{\bibfnamefont{G.}~\bibnamefont{Ford}} \bibnamefont{and}
  \bibinfo{author}{\bibfnamefont{M.}~\bibnamefont{Kac}},
  \bibinfo{journal}{J.~Stat.~Phys.} \textbf{\bibinfo{volume}{46}},
  \bibinfo{pages}{803} (\bibinfo{year}{1987}).

\bibitem[{\citenamefont{Ford and Lewis}(1986)}]{FordLewis86}
\bibinfo{author}{\bibfnamefont{G.}~\bibnamefont{Ford}} \bibnamefont{and}
  \bibinfo{author}{\bibfnamefont{J.}~\bibnamefont{Lewis}}, in
  \emph{\bibinfo{booktitle}{Probability, Stochastics, and Number Theory}}
  (\bibinfo{publisher}{Academic Press}, \bibinfo{year}{1986}),
  vol.~\bibinfo{volume}{9} of \emph{\bibinfo{series}{Advances in Mathematics
  Supplemental Studies}}, p. \bibinfo{pages}{169}.

\bibitem[{\citenamefont{Grabert et~al.}(1988)\citenamefont{Grabert, Schramm,
  and Ingold}}]{GrabertIngold88}
\bibinfo{author}{\bibfnamefont{H.}~\bibnamefont{Grabert}},
  \bibinfo{author}{\bibfnamefont{P.}~\bibnamefont{Schramm}}, \bibnamefont{and}
  \bibinfo{author}{\bibfnamefont{G.-L.} \bibnamefont{Ingold}},
  \bibinfo{journal}{Phys. Rep.} \textbf{\bibinfo{volume}{168}},
  \bibinfo{pages}{115} (\bibinfo{year}{1988}).

\bibitem[{\citenamefont{Strunz and Haake}(2003)}]{StrunzHaake03}
\bibinfo{author}{\bibfnamefont{W.}~\bibnamefont{Strunz}} \bibnamefont{and}
  \bibinfo{author}{\bibfnamefont{F.}~\bibnamefont{Haake}},
  \bibinfo{journal}{Phys. Rev. A} \textbf{\bibinfo{volume}{67}}, \bibinfo{pages}{022102}
  (\bibinfo{year}{2003}).

\bibitem[{\citenamefont{Hakim and Ambegaokar}(1985)}]{HakimAmbegaokar85}
\bibinfo{author}{\bibfnamefont{V.}~\bibnamefont{Hakim}} \bibnamefont{and}
  \bibinfo{author}{\bibfnamefont{V.}~\bibnamefont{Ambegaokar}},
  \bibinfo{journal}{Phys. Rev. A} \textbf{\bibinfo{volume}{32}},
  \bibinfo{pages}{423} (\bibinfo{year}{1985}).

\bibitem[{\citenamefont{Feynman and Hibbs}(1965)}]{FeynmanHibbs65}
\bibinfo{author}{\bibfnamefont{R.}~\bibnamefont{Feynman}} \bibnamefont{and}
  \bibinfo{author}{\bibfnamefont{A.}~\bibnamefont{Hibbs}},
  \emph{\bibinfo{title}{Quantum mechanics and path integrals}}
  (\bibinfo{publisher}{McGraw-Hill}, \bibinfo{year}{1965}).

\bibitem[{\citenamefont{Caldeira and Leggett}(1983)}]{CaldeiraLeggett83}
\bibinfo{author}{\bibfnamefont{A.}~\bibnamefont{Caldeira}} \bibnamefont{and}
  \bibinfo{author}{\bibfnamefont{A.}~\bibnamefont{Leggett}},
  \bibinfo{journal}{Physica A} \textbf{\bibinfo{volume}{121}},
  \bibinfo{pages}{587} (\bibinfo{year}{1983}).

\bibitem[{\citenamefont{Zurek}(1991)}]{Zurek91}
\bibinfo{author}{\bibfnamefont{W.}~\bibnamefont{Zurek}},
  \bibinfo{journal}{Physics Today} \textbf{\bibinfo{volume}{44(10)}},
  \bibinfo{pages}{36} (\bibinfo{year}{1991}). See also Ref.~\olcite{ambeg93}.

\bibitem[{\citenamefont{Ford and O'Connell}(2001)}]{FordOConnell01c}
\bibinfo{author}{\bibfnamefont{G.}~\bibnamefont{Ford}} \bibnamefont{and} 
  \bibinfo{author}{\bibfnamefont{R.}~\bibnamefont{O'Connell}},
  \bibinfo{journal}{Am.J.Phys} \textbf{\bibinfo{volume}{70}},
  \bibinfo{pages}{319} (\bibinfo{year}{2001}).

\bibitem[{\citenamefont{Murakami, Ford and O'Connell}(2003)}]{MurakamiOConnell03}
  \bibinfo{author}{\bibfnamefont{M.}~\bibnamefont{Murakami}},
  \bibinfo{author}{\bibfnamefont{G.}~\bibnamefont{Ford}} \bibnamefont{and}
  \bibinfo{author}{\bibfnamefont{R.}~\bibnamefont{O'Connell}},
  \bibinfo{journal}{Laser Physics} \textbf{\bibinfo{volume}{13}},
  \bibinfo{pages}{180} (\bibinfo{year}{2003}).


\end{thebibliography}
\end{document}